\documentclass[twocolumn]{jpsj3}
\usepackage[T1]{fontenc}
\usepackage[latin9]{inputenc}
\usepackage{graphicx}
\usepackage{amsmath,amsbsy,amssymb}
\usepackage{bm}
\usepackage{color}

\title{Exact dynamics of charge fluctuations in the multichannel interacting resonant level model}

\author{Annam\'aria \textsc{Kiss}\thanks{E-mail address: kiss.annamaria@wigner.mta.hu}$^{1,2}$, Yoshio \textsc{Kuramoto}$^{3,4}$, and Junya \textsc{Otsuki}$^{3}$}

\inst{
$^1$Institute for Solid State Physics and Optics, Wigner Research Centre for Physics, Hungarian Academy of Sciences, P. O. B. 49, H-1525 Budapest, Hungary
\\
$^2$BME-MTA Exotic Quantum Phases "Lend\"ulet" Group, Institute of Physics, Budapest University of Technology and Economics, H-1521 Budapest, Hungary\\
$^3$Department of Physics, Tohoku University, Sendai, 980-8578, Japan\\
$^4$Institute of Material Structure Science, KEK, Tsukuba, Ibaraki 305-0801, Japan
}

\recdate{\today}

\abst{
A modified version of the spinless Anderson model is studied by means of the continuous-time quantum Monte Carlo method.  This study is motivated by the peculiar
heavy-fermion behavior observed in certain Samarium compounds, which is insensitive to 
magnetic field.
The model involves $M$ channels for conduction electrons, all of which interact with 
local $f$ electron via the Coulomb repulsion $U_{fc}$, while only one channel has 
hybridization with the local state.
The effective hybridization is reduced by the Anderson orthogonality effect, and a quantum critical point occurs with increasing $M$ and/or increasing $U_{fc}$.
The numerical results at finite temperature of the local charge susceptibility 
are well fitted by a simple scaling theory for all $M$.  However, the 
single-particle spectrum is described by a double Lorentzian for $M>1$, in contrast with the single Lorentzian with $M=1$.
A quasi-particle perturbation theory is presented that reproduces the quantum critical point for large $M$.  The quasi-particle theory gives not only the renormalized energy scale, but its extrapolation toward higher energies being consistent with the double Lorentzian spectrum.
}

\kword{charge Kondo effect, continuous-time quantum Monte Carlo method, quasi-particle perturbation theory, thermodynamic and dynamic properties}

\begin{document}
\maketitle

\section{Introduction}

Recently, peculiar heavy-fermion behavior has attracted attention in certain Samarium compounds with large specific heat coefficient $\gamma$ which is insensitive to external magnetic field. 
For example, the filled skutterudite compound SmOs$_{4}$Sb$_{12}$ has $\gamma \sim 0.8 {\rm J/(K^2\cdot mol)}$ even though it is mixed valent\cite{sanada-2005}.  Similar behavior has been found in systems such as
SmPt$_{4}$Ge$_{12}$\cite{Gumeniuk} and SmT$_{2}$Al$_{20}$ with T=Ti,V,Cr,Ta\cite{sakai,Higashinaka-1,Higashinaka-2}.
The resistivity of SmT$_{2}$Al$_{20}$ shows clear Kondo-like logarithmic temperature dependence, which however is insensitive to external magnetic field\cite{sakai}.  
It has been suspected that charge degrees of freedom is 
responsible for the heavy mass
because of the field-insensitivity, in striking contrast to ordinary Kondo effect which is  
sensitive to magnetic field.

Motivated by these experimental observations we search for a charge fluctuation mechanism that gives rise to energy scale much smaller than bare hybridization. 
As the simplest attempt,
we study the (spinless) multichannel interacting resonant level model (MIRLM) by means of the continuous-time quantum Monte Carlo method, which starts from the Anderson model involving the hybridization ($V$) of the local charge with only one of the conduction electron orbitals, 
but includes additional Coulomb interaction ($U_{fc}$) felt by all conduction orbitals with the local $f$ state.
The Hamiltonian of this model is written as
\begin{align}
{\cal H} &=  {\cal H}_{c} + {\cal H}_{f} + {\cal H}_{hyb} + {\cal H}_{fc} \nonumber\\
&= \sum_{\boldsymbol{k}} \sum_{\ell=0}^{M-1} \varepsilon_{\boldsymbol{k}} c^{\dag}_{\boldsymbol{k}, \ell}c_{\boldsymbol{k},\ell} + \varepsilon_{f}   f^{\dag}f +
 V(f^{\dag}c_{0} + c^{\dag}_{0}f)\nonumber\\
 &+ U_{fc} \sum_{\ell=0}^{M-1} \left( f^{\dag}f - \frac{1}{2} \right) \left( c^{\dag}_{\ell}c_{\ell} - \frac{1}{2} \right),
\label{eq-ham-0}
\end{align}
where $M$ is the number of the conduction electron channels, $c_{\boldsymbol{k}, \ell}$ ($c^{\dag}_{\boldsymbol{k}, \ell}$) is the annihilation (creation) operator of the Bloch state $\boldsymbol{k}$ in the $\ell$th channel, and 
$c_{\ell}=N^{-1/2}\sum_{\boldsymbol{k}} c_{\boldsymbol{k},\ell}$ with $N$ being the number of sites.

The model given by Eq.~(\ref{eq-ham-0}) leads to rich physics under finite values of the Coulomb interaction $U_{fc}$ with more than one conduction channels ($M>1$). This is due to the increasing dominance of the Anderson orthogonality effect arising from the screening channels over the exciton enhancement coming from the hybridizing channel. Namely, the presence of multiple channels of conduction electrons leads to non-trivial low-energy renormalization of bare $V$.

In addition to the motivation provided by the peculiar heavy-fermion state of certain Samarium compounds, such quantum impurity models play important role to understand many-body phenomena realized in single artificial atoms, or quantum dots, with multiple levels interacting with a Fermi sea and encountering further local interactions. 

The single-channel version of the model has been extensively studied by many authors with various analytic methods including bosonization\cite{schlottmann-1}, Bethe ansatz\cite{wiegmann}, Anderson-Yuval mapping to Coulomb gas\cite{AY} and perturbative renormalization\cite{schlottmann-2,schlottmann-3}.
However, much less is known about the multichannel version of the model.
Although the original interacting resonant level model has been extended to multiple channels by perturbative renormalization\cite{giamarchi} and also been studied by numerical renormalization group\cite{borda-2007, borda-2008}, these works did not discuss the dynamics under finite hybridization and at finite temperatures.

In a previous paper\cite{IRLM2013}, we already studied numerically the single-channel ($M=1$) version of the model in the negative $U_{fc}$ range 
and provided quantitative information about the dynamics at finite temperatures. 
In this paper we extend the study for the positive $U_{fc}$ range and for the presence of multiple conduction channels.  
By means of the continuous-time quantum Monte Carlo method we 
investigate both thermodynamic and dynamic properties of MIRLM 
in a wide range of the Coulomb interaction $U_{fc}$.
Especially, we are interested in the applicability and accuracy of perturbative approaches. 
 
This paper is organized as follows.
In Section~\ref{sec-analytic_approach} we derive the renormalized hybridization for the multi-channel case by means of perturbative renormalization within a simple scaling theory.
In Section~\ref{sec-numerical_approach} the continuous-time quantum Monte Carlo algorithm is formulated.
The numerically obtained static and dynamic properties are presented in Sections~\ref{sec-static} and \ref{sec-dynamic}.
We construct a quasi-particle perturbation theory in Section~\ref{sec-discussion} in order to
better understand the numerical results. Finally, Section~\ref{sec-summary} is devoted to the  summary of this paper.

\section{Perturbative Renormalization Approach}\label{sec-analytic_approach}

\subsection{Scaling energy with multiple conduction channels}

The multichannel version of the interacting resonant level model
was first introduced in Ref.\citen{giamarchi}, which mapped the MIRLM to an anisotropic Kondo-like model and derived the scaling equations of the corresponding Kondo model.
Here, we follow a different way. Namely, we extend the method that we used in our previous work\cite{IRLM2013} to the case of multi channels for the conduction electrons to obtain the hybridization renormalized by the Coulomb interaction.
In that work we derived the effective hybridization for the single-channel case by using the effective Hamiltonian method\cite{kuramoto-1998} to take account of the simultaneous effect of the hybridization $V$ and Coulomb interaction $U_{fc}$. 
If we include multiple conduction channels, the expression for the renormalized hybridization is modified as
\begin{eqnarray}
V^{\prime} = V \left(1 - u \frac{\delta D}{D} + \frac{1}{2}u^2M \frac{\delta D}{D} \right),\label{renV0}
\end{eqnarray}
where $D$ is the band cutoff and we introduced a dimensionless coupling constant $u=\rho_{0} U_{fc}$ with $\rho_0$ being the density of states.
The 
factor $M$
accounts for the 
closed conduction-electron loops that represent the Anderson orthogonality effect.
Following the same procedure as 
in Ref.~\citen{IRLM2013}, the effective hybridization is finally obtained as
\begin{eqnarray}
V^{\ast} = V \left(\frac{\Delta_{0}}{D} \right)^{(-u+Mu^2/2)/(1+2u-Mu^2)},
\label{eq-renV}
\end{eqnarray} 
where $\Delta_{0} \equiv \pi \rho_{0}V^2$ is introduced.
Correspondingly, $V^{\ast}$ defines the scaling energy $\Delta^{\ast}_{sc}=\pi \rho_{0}(V^{\ast})^2$, and we express from Eq.~(\ref{eq-renV}) as
\begin{eqnarray}
{\rm ln} \frac{\Delta^{\ast}_{sc}}{\Delta_{0}} = -\eta {\rm ln} \frac{\Delta^{\ast}_{sc}}{D} \label{eq-eta}
\end{eqnarray}
with 
\begin{equation}
\eta \equiv u(2-Mu) 
\label{eq-eta2}.
\end{equation}
Coefficient $\eta$ 
controls $\Delta^{\ast}_{sc}$, and therefore the effective hybridization $V^{\ast}$. Namely, the exciton effect ($\sim u$) coming from the single hybridizing channel enhances the hybridization with increasing Coulomb interaction, while the Anderson orthogonality effect ($\sim -Mu^2$) from  
all channels acts against the exciton effect by blocking the hybridization. 
As a result, the competition of these two effects renormalizes the hybridization at low energies in a non-trivial way.

The renormalized hybridization obtained in Eq.~(\ref{eq-renV})  is valid only for 
small values of the bare parameters $V$ and $U_{fc}$ because of the perturbative renormalization treatment. 
Making use of the analogy with the x-ray threshold problem\cite{nozieres-1969} by considering finite $U_{fc}$ but neglecting its interference effect with the infinitesimal hybridization, an associated phase shift can be introduced\cite{schlottmann-2,schlottmann-3} as
\begin{eqnarray}
\tilde{u} = \frac{\delta_{U}}{\pi} = \frac{2}{\pi} {\rm tan}^{-1}(\pi \rho_{0}U_{fc}/2)
\label{eq-phasesh}
\end{eqnarray}
in place of $u = \rho_{0}U_{fc}$ to account for the multiple scattering by $U_{fc}$ to infinite order.

We have found in Ref.~\citen{IRLM2013} that the phase shift picture accounts quite accurately for the effective hybridization in the negative $U_{fc}$ range in the single-channel case.

\subsection{Scaling energy at finite temperature}

Let us first quote Schlottmann's extension\cite{schlottmann-3} of the energy scale $\Delta^{\ast}_{sc}$ to finite temperatures ($T$) and frequencies ($\varepsilon$):
\begin{align}
{\rm ln} \frac{\Delta^{\ast}_{sc}(\varepsilon,T)}{\Delta_{0} } &= \eta \left[ {\rm log} \left(\frac{D}{2\pi T} \right)
\right. \nonumber\\
 -& \left. {\rm Re}\, \psi \left(\frac{1}{2} + \frac{\Delta^{\ast}_{sc}(\varepsilon,T)}{2\pi T} -i \frac{\varepsilon}{2\pi T} \right) \right],  
\label{eq-schomd}
\end{align}
where $\psi$ is the digamma function. 
Regarding the energy dependence of $\Delta^{\ast}_{sc}(\varepsilon,T)$, the above expression is 
an interpolation formula between $\Delta^{\ast}_{sc}$ given in Eq.~(\ref{eq-eta}) at $\varepsilon=0$, and the result of the x-ray edge problem $\Delta_{0}(|\varepsilon|/D)^{\eta}$ for  $|\varepsilon| \gg \Delta^{\ast}_{sc}$.
Based on the idea that the interacting resonant level problem is still described by a resonance with the renormalized width $\Delta^{\ast}_{sc}(\varepsilon,T)$ under finite $U_{fc}$ instead of the bare $\Delta_{0}$, Schlottmann obtained the charge susceptibility as\cite{schlottmann-3}
\begin{align}
\chi^{\ast}_{c}(\varepsilon,T) &= - \frac{2\Delta^{\ast}_{sc}(T)}{\pi \varepsilon} \frac{1}{(\varepsilon + i 2 \Delta^{\ast}_{sc}(T))}
 \nonumber\\
\times&
\left[
\psi \left( \frac{1}{2} + \frac{\Delta^{\ast}_{sc}(T)}{2\pi T} - i \frac{\varepsilon}{2\pi T} \right)
\right. \nonumber\\
 -& \left.
 \psi \left( \frac{1}{2} + \frac{\Delta^{\ast}_{sc}(T)}{2\pi T} \right)
 \right]
 \label{eq-chischl1}
\end{align}
by taking the convolution of two simple resonances associated to the $f$-electron Green's function.
The static component is given as $\chi^{\ast}_0 \equiv \chi^{\ast}_{c}(\varepsilon=0)$, which is obtained as
\begin{eqnarray}
\chi^{\ast}_{0}(T=0) = \frac{1}{\pi \Delta^{\ast}_{sc}} \label{eq-chiT0}
\end{eqnarray}
at zero temperature from Eq.~(\ref{eq-chischl1}).

Equation~(\ref{eq-chiT0}) expresses that the charge susceptibility is scaled with the single scaling energy $\Delta^{\ast}_{sc}$. This can be understood by recalling 
the Ward identity\cite{schlottmann-3, zawadowski, LW1}, which is a constraint for the correlation functions dictated by conservation laws present in a model.
In the case of the resonant level problem the total number of the local and conduction electrons, i.e. the charge is conserved, which leads to a relation between the vertex corrections and self energies.\cite{schlottmann-3} 
As it was shown by {\sl Schlottmann} in Ref.~\citen{schlottmann-3}, a consequence of the Ward identity for the resonant level problem is that the vertex corrections and self-energy cancel each other in quasi-particles, and thus the quasi-particle density of states $\rho_{f}(\varepsilon)$ is completely determined by the scaling energy $\Delta^{\ast}_{sc}$ as
\begin{eqnarray}
\rho_{f} (\varepsilon) = \frac{1}{\pi} \frac{\Delta^{\ast}_{sc}}{\left( (\varepsilon - \varepsilon_{f})^2 + (\Delta^{\ast}_{sc})^2 \right)}.
\label{eq-ron}
\end{eqnarray}
By further use of the Ward identity, the following relation was obtained\cite{schlottmann-3} between the charge susceptibility and specific heat coefficient $\gamma$ at zero temperature:
\begin{eqnarray}
3\gamma/\pi^2  = \chi_{0}^{\ast} = \rho_{f}(i\delta) = \frac{1}{\pi \Delta^{\ast}_{sc}}
, \label{eq-chicWard2n}
\end{eqnarray}
i.e. $\chi_{0}^{\ast}$ is entirely determined by non-interacting quasiparticles through the resonant level given in Eq.~(\ref{eq-ron}). We can apply the argument above also to the case with $M>1$.

\subsection{Quantum critical points}

The competition of the exciton effect with the Anderson orthogonality effect reflected in coefficient $\eta$ given in Eq.~(\ref{eq-eta2}) drives the system toward a quantum critical point at about $u \sim 2/M$.
To be more precise, 
we write the exponent $x(u,M)$ in Eq.~(\ref{eq-renV}) as
\begin{align}
x(u,M) &\equiv \frac{-u+Mu^2/2}{1+2u-Mu^2} \nonumber\\
 &= -\frac{1}{2} - \frac{1}{2M (u-u_{-})(u-u_{+})}
\end{align}
with
\begin{eqnarray}
u_{\pm}(M) = \frac{1}{M}(1 \pm \sqrt{1+M}).\label{eq-upm}
\end{eqnarray}
When $u$ approaches to  $u_{\pm}$, the exponent $x(u,M)$ diverges to positive infinity, which means that the effective hybridization $V^{\ast}$ vanishes, i.e. the local $f$ charge becomes decoupled from the conduction electrons. The point $V^{\ast}=0$ is equivalent with the ferromagnetic Kondo fixed point  with degeneracy between the empty and occupied $f$ states.
We note that the value $|u|_{\pm} \sim {\cal O}(1/\sqrt{M})$ obtained in Eq.~(\ref{eq-upm}) for $M \gg 1$ is within the range of the perturbative treatment, while $|u|_{\pm} \sim {\cal O}(1)$ for $M=1$ might be artificial.

Taking the positive $U_{fc}$ range, the perturbative treatment with $u=\rho_{0} U_{fc}$ predicts a quantum critical point at $u=u_{+}=1+\sqrt{2}$ even for the single-channel case. On the other hand, in the phase shift picture, 
the condition $1+2\tilde{u}-M\tilde{u}^2=0$ of vanishing hybridization requires $M\ge 3$
with $\tilde{u}=1$ corresponding to the maximum phase shift $\delta_{U}=\pi$.
The study of MIRLM by means of numerical renormalization group method\cite{borda-2007, borda-2008} found a saturation of the renormalized hybridization for $M=1$ in the positive $U_{fc}$ range,
and vanishing $V^{\ast}$ with increasing $U_{fc}$ for $M \ge 3$.
However, it 
remains to see what happens for $M=2$.
Although the NRG study found
a suppression of the renormalized hybridization for $M=2$ by increasing $U_{fc}$, 
it cannot be decided explicitly from the data whether it reaches zero or not.

\section{Continuous-Time Quantum Monte Carlo Approach}\label{sec-numerical_approach}

In this section, we analyze the MIRLM using the continuous-time quantum Monte Carlo method~\cite{gull-2011}.
In the previous paper\cite{IRLM2013}, we presented an algorithm for the single-channel case, $M=1$, based on an expansion with respect to $V$ and $U_{fc}$. 
The advantage of the double-expansion algorithm compared with an ordinary weak-coupling expansion with respect to $U_{fc}$ is that the computational cost increases only linearly as $M$ is increased, yielding efficient calculations for large $M$.
Since the extension to $M>1$ is straightforward, we only briefly describe difference
from the case with $M=1$ in the following.

\begin{figure}
\centering
\includegraphics[width=0.85\hsize]{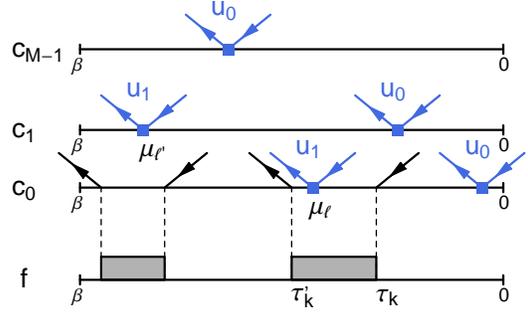}
\caption{(Color online) An example of the Monte Carlo configuration.}
\label{fig-config}
\end{figure}

The Monte Carlo configuration is expressed schematically in Fig.~\ref{fig-config}. 
Here
the imaginary time sequence
${\boldsymbol \tau}=\left\{ \tau_{1}, \tau_{1}^{\prime}, \dots,  \tau_{q}, \tau_{q}^{\prime}  \right\}$
defines the $V$-expansion process.
The interaction $U_{fc}$ is described by a time-dependent potential which fluctuates between $u_0$ and $u_1$ depending on the occupation of the $f$ state.\cite{IRLM2013}
The existence of the extra channels $M \geq 1$ modifies the weight
$W_{c}$, which is factorized as
\begin{eqnarray}
W_{c}({\boldsymbol \tau},{\boldsymbol \mu}_0,\cdots,{\boldsymbol \mu}_{M-1}) = {\rm det}D_{0}({\boldsymbol \tau},{\boldsymbol \mu}_{0}) \prod_{\ell=1}^{M-1} {\rm det} D_{\ell}({\boldsymbol \mu}_{\ell}),
\label{eq-Wc}
\end{eqnarray}
where ${\boldsymbol \mu_l}$ denotes a set of imaginary times at which scattering takes place.
$D_{\ell}$ is a $m_l \times m_l$ matrix consisting of the bare Green's functions connecting two time points in ${\boldsymbol \mu_l}$.

\begin{figure}
\centering
\includegraphics[width=0.95\hsize]{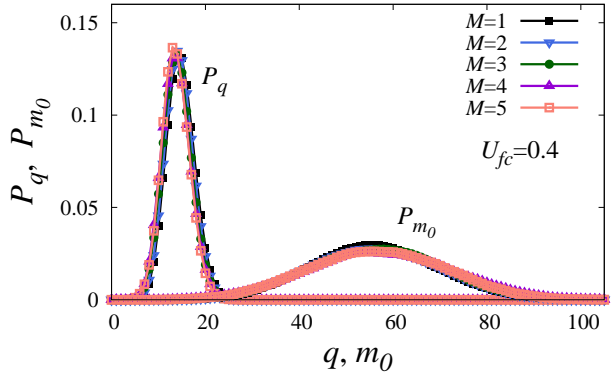}
\caption{(Color online) Histograms $P_{q}$ and $P_{m_0}$ for different values of channel number $M$. Here $q$ denotes the $V$-expansion order and $m_0$ the $U_{fc}$-expansion order in the 0th channel.
The parameter values are chosen as $U_{fc}=0.4$, $V=0.2$ and $\beta=200$.}
\label{fig-probability}
\end{figure}

The last factor $\det D_{\ell}$ requires an extra cost compared to the case of $M=1$. 
The important point is that the expansion order $m_{\ell}$ for each channel is not changed so much when $M$ is increased, as confirmed from the histogram in Fig.~\ref{fig-probability}. 
It means that the size of the matrix $D_{\ell}$ is almost independent of $M$ and therefore, 
the computational cost increases only linearly against $M$.
In the weak-coupling algorithm, on the other hand, one computes the determinant of a $\sum_l m_l \times \sum_l m_l$ matrix for the $f$ state. It means that the matrix size is proportional to $M$ and hence, the cost increases according to ${\cal O}(M^2)$.

The single-particle Green's function consists of three components, $G_f$, $G_c$ and $G_{fc}$.
The relation between each component and the self-energy is summarized in Appendix~A.
In numerical calculations, we use the constant density of states $\rho_0=1/(2D)$ with the band cutoff $D=1$ for all channels.
We restrict ourselves to the particle-hole symmetric case, $\varepsilon_{f}=0$. 
Spectra in the real-frequency domain are obtained by analytic continuation $i\varepsilon_{n} \rightarrow \varepsilon + i\delta$ using the Pad\'e approximation.
We imposed the condition for the particle-hole symmetry,
${\rm Re}\, G_{f}(i\varepsilon_{n}) = {\rm Re}\, G_{c}(i\varepsilon_{n}) = {\rm Im}\, G_{fc}(i\varepsilon_{n})=0$, 
to improve accuracy.

\section{Static Properties}\label{sec-static}

We first discuss the static charge susceptibility $\chi_0(T)$ to check applicability of the scaling theory in Section~\ref{sec-analytic_approach}.
By taking the limit $\varepsilon=0$ in Eq.~(\ref{eq-chischl1}), we obtain the analytic expression $\chi_0^*(T)$ for the static charge susceptibility as
\begin{eqnarray}
\chi^{\ast}_{0}(T) 
= \frac{1}{2\pi^2 T} \psi \left( \frac{1}{2} + \frac{\Delta^{\ast}_{sc}(T)}{2\pi T} \right).\label{eq-schchi0}
\end{eqnarray}
In comparing this expression with our numerical results, we should take care of influence of the band cutoff $D$: the analytic expression was derived in the limit $D\to \infty$, while the simulation is performed with a finite value of $D$.
We introduce a correction factor $\alpha=1-2\Delta/(\pi D)$ and replace $\chi_0^*$ with $\alpha \chi_0^*$ to take the influence of finite $D$ into account\cite{otsuki-2007}.

\begin{figure}
\centering
\includegraphics[width=0.95\hsize]{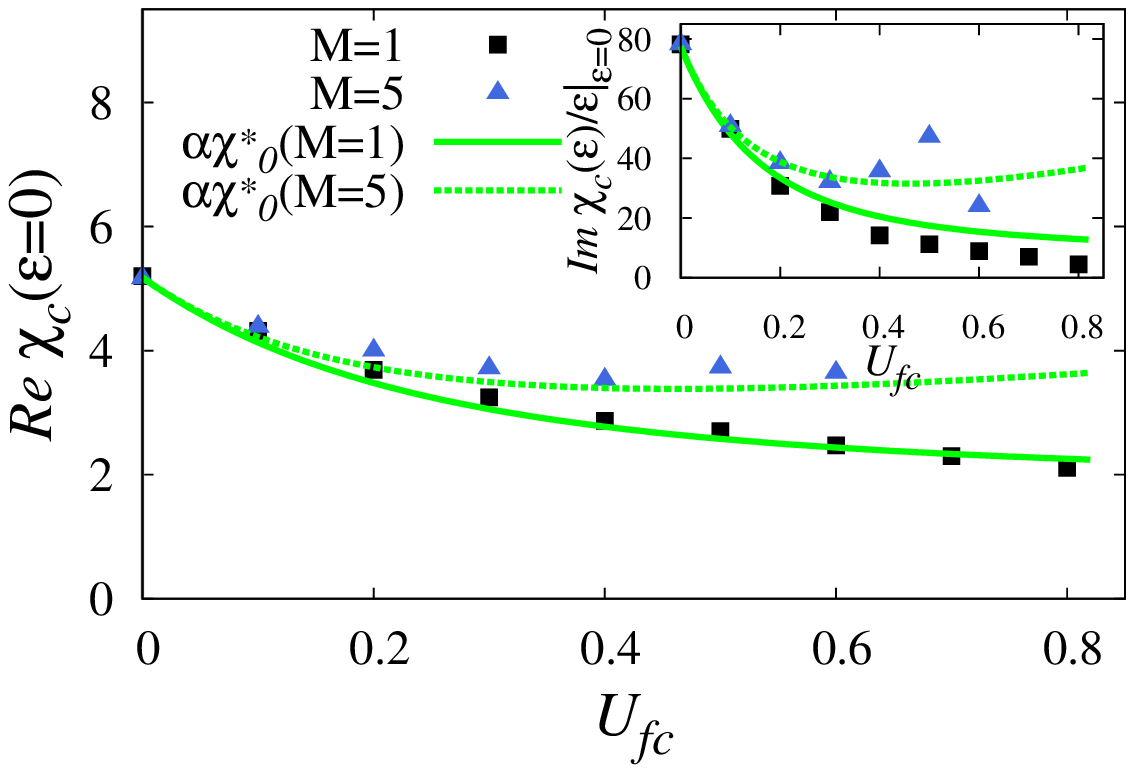}
\includegraphics[width=0.95\hsize]{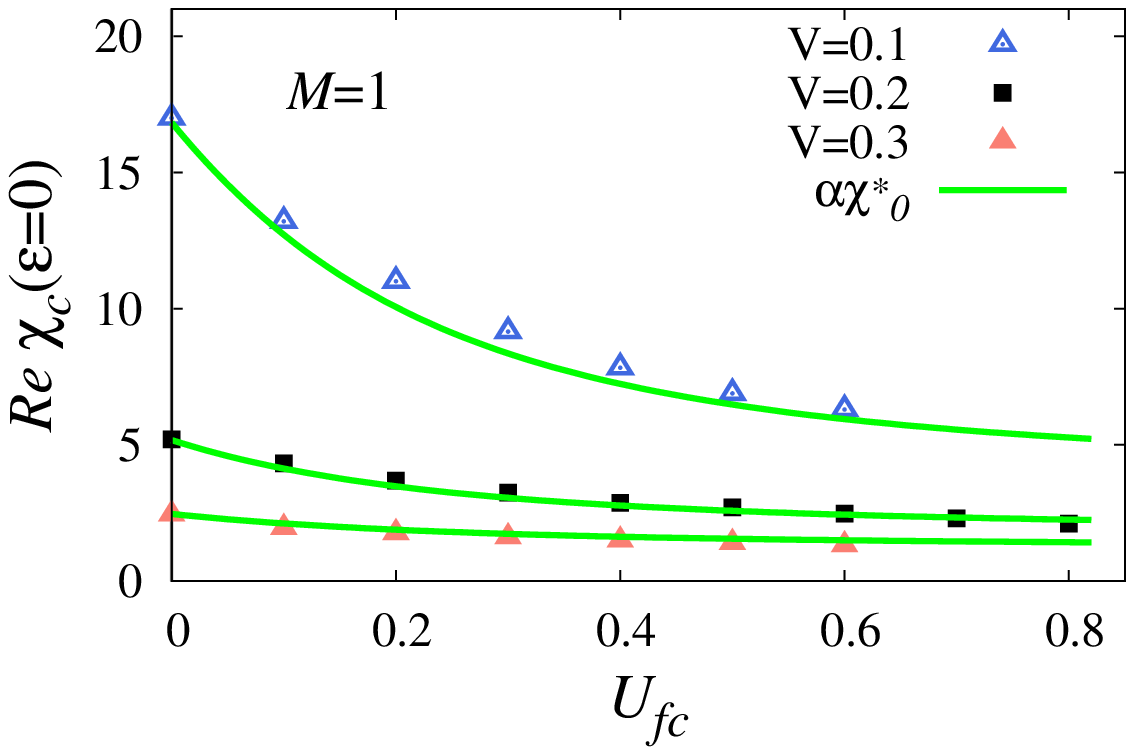}
\caption{(Color online) Numerical result for static charge susceptibility as a function of $U_{fc}$ for $\beta=200$. {\sl Top:} Comparison between $M=1$ and $M=5$ with $V=0.2$ fixed.
{\sl Bottom:} Comparison of different values of $V$ for $M=1$. 
The {\sl inset} shows the $U_{fc}$-dependence of the zero-energy limit of the imaginary part of the charge susceptibility.
Schlottmann's susceptibility $\chi^{\ast}_{0}$ is shown by {\sl solid green lines}.}
\label{fig-charge-susceptibility}
\end{figure}

Top part of Fig.~\ref{fig-charge-susceptibility} shows comparison between the numerical results $\chi_0$ and the analytic expression $\alpha \chi_0^*$ for $M=1$ and $M=5$.
The analytic expression turns out to give an excellent fit of the numerical data in the wide range of $U_{fc}$.
Thus, we conclude that $\chi_0$ is scaled with a single parameter $\Delta^{\ast}_{sc}$
as it was discussed in Section~2.
We also make a comparison of the imaginary part of the charge susceptibility, 
${\rm Im}\, \chi_{c}(\varepsilon)/\varepsilon|_{\varepsilon=0}$,
in the inset of Fig.~\ref{fig-charge-susceptibility}.
It can be seen that the agreement is good for moderate values of $U_{fc}$. 
However, a systematic deviation is observed for a large value of $U_{fc}$ with $M=1$.
It might be caused by the neglect of the energy dependence of $\Delta^{\ast}_{sc}$ in Schlottmann's formula~(\ref{eq-chischl1}).

Finally, we comment on the hybridization-dependence of the charge susceptibility.
Although arbitrary value of the Coulomb interaction $U_{fc}$ can be handled by the associated phase shift introduced in Eq.~(\ref{eq-phasesh}), 
Schlottmann's formula~(\ref{eq-schchi0}) is valid only for small values of the bare hybridization.
Thus, it is interesting to check the validity of 
$\chi_0^{\ast}$
as the value of the bare hybridization $V$ is increased.
Bottom part of Fig.~\ref{fig-charge-susceptibility}
shows the charge susceptibility obtained for different values of $V$ together with 
$\alpha \chi_0^{\ast}$. 
Surprisingly, the agreement is very good even for the largest value of $V=0.3$, which is already comparable to the half-bandwidth $D=1$ used in the simulations.

\section{Dynamic Properties}\label{sec-dynamic}

\subsection{Green's function}\label{sec-gf}

\begin{figure}
\centering
\includegraphics[width=0.95\hsize]{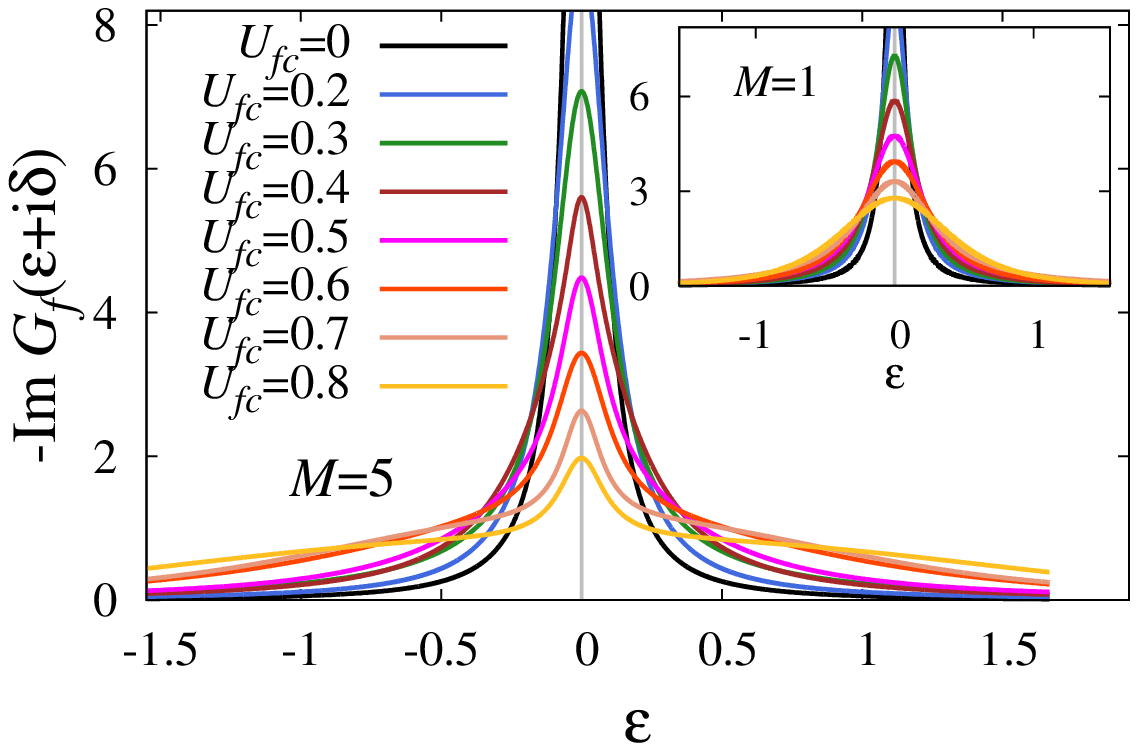}
\includegraphics[width=0.95\hsize]{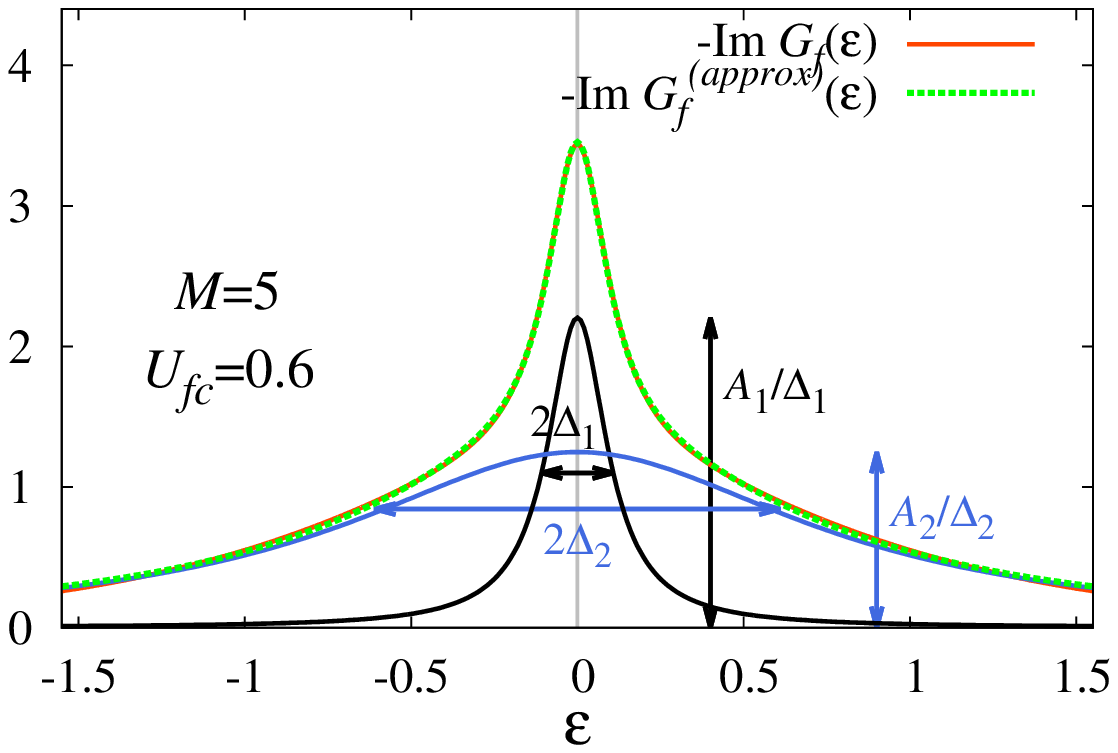}
\caption{(Color online) {\sl Top:} Energy dependence of the imaginary part of the $f$-electron Green's function for channel numbers $M=5$ ({\sl main panel}) and $M=1$ ({\sl inset}). 
{\sl Bottom:} Fit of the spectral function by the sum of two Lorentzians at $U_{fc}=0.6$ for $M=5$.
The parameter values are chosen as $V=0.2$, $\beta=200$.}
\label{fig-ftdos}
\end{figure}

Top part of Fig.~\ref{fig-ftdos} shows the single-particle excitation spectrum, $-{\rm Im}G_f(\varepsilon+i\delta)$, with $M=5$ for several values of $U_{fc}$.
As $U_{fc}$ is increased, a distinct deviation is observed compared with the non-interacting lineshape, i.e., the Lorentzian; the spectrum exhibits a high-energy tail in addition to a sharp peak around $\varepsilon=0$.
Since the high-energy tail is not observed in $M=1$ as shown in the {\sl inset} of Fig.~\ref{fig-ftdos}, it is due to the extra screening channels, $\ell \neq0$, without hybridization.

By analysis of the numerical data, we found that the peculiar spectra observed for $M>1$ can be well approximated by a sum of two Lorentzians:
\begin{eqnarray}
G_{f}^{\rm (approx)}(z)  \equiv  \frac{A_{1}}{z + i \Delta_{1}} + \frac{A_{2}}{z + i \Delta_{2}}.
\label{eq-zfdosLorentz}
\end{eqnarray}
Here, the first term describes the $f$ level peaked at $\varepsilon_f=0$, 
while the second term gives an account of the high-energy tail. 
A fitting result with $U_{fc}=0.6$ for $M=5$ is shown in the bottom part of Fig.~\ref{fig-ftdos}.
An excellent agreement is confirmed.
The function in Eq.~(\ref{eq-zfdosLorentz}) is reduced to a single Lorentzian when $A_2=0$, which should take place for $U_{fc}=0$ or $M=1$.

\begin{figure}
\centering
\includegraphics[width=0.95\hsize]{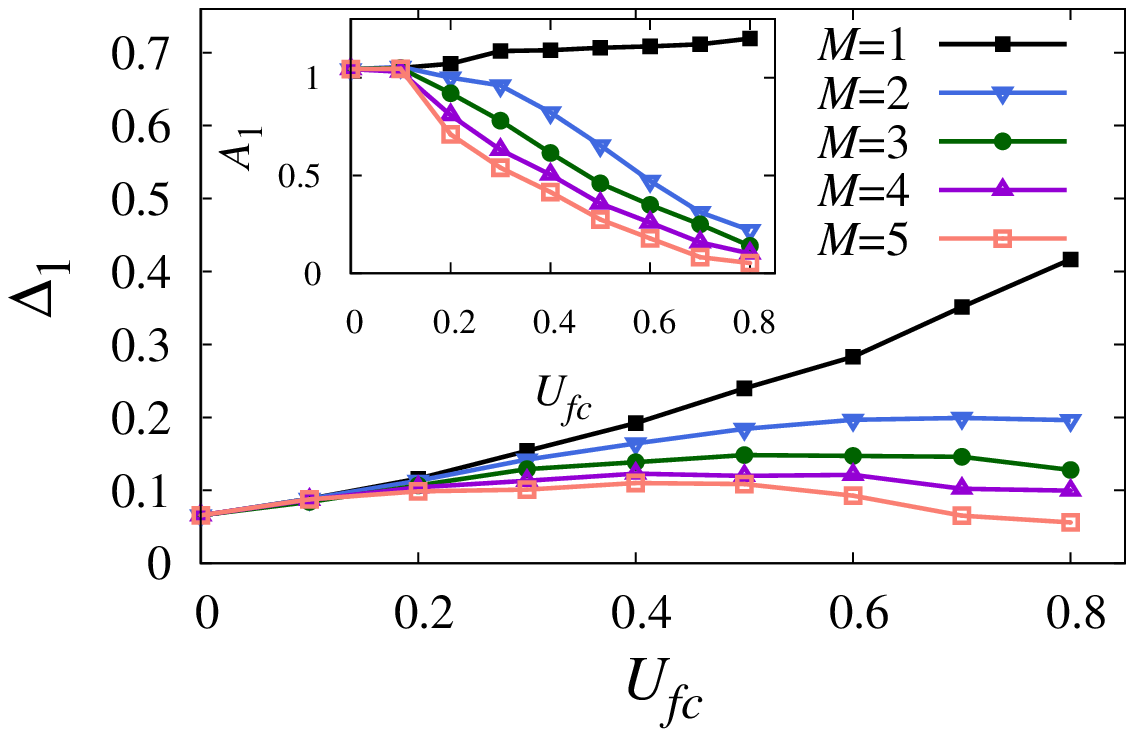}
\includegraphics[width=0.95\hsize]{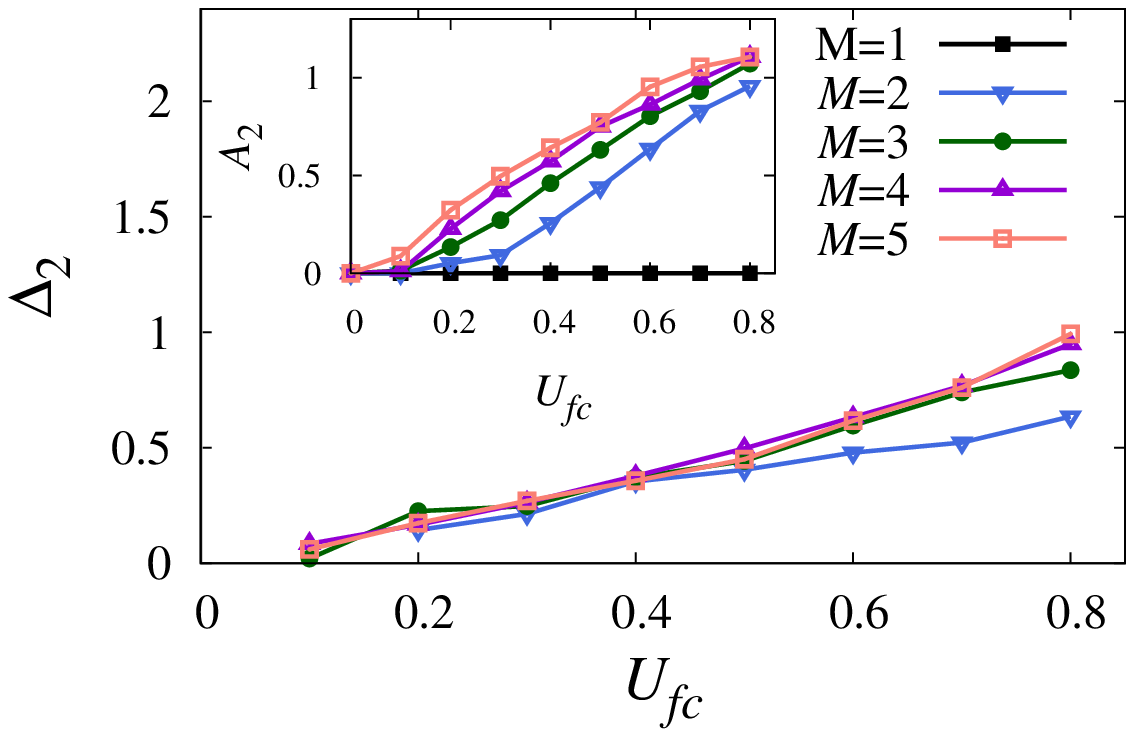}
\caption{(Color online) Fitting result for the parameters $\Delta_{1}$, $A_{1}$ ({\sl top}) and $\Delta_{2}$, $A_{2}$ ({\sl bottom})  as a function of $U_{fc}$ with increasing channel number $M$.
The parameter values are chosen as $V=0.2$, $\beta=200$.}
\label{fig-parameters}
\end{figure}

Numerical results for the fitting parameters $\Delta_1$, $\Delta_2$, $A_1$ and $A_2$ are plotted in Fig.~\ref{fig-parameters} as a function of $U_{fc}$ for several values of $M$.
In these calculations, we have worked in the Matsubara frequency domain to avoid 
inaccuracy
caused by analytic continuations.
For $M=1$, we obtain a single Lorentzian, $A_1 \approx 1$ and $A_2=0$, as expected.
The 
improper result, $A_1>1$, is due to influence of the finite bandwidth used in the simulations. 
For $M>1$, the weight $A_1$ is transfered to $A_2$ as $U_{fc}$ increases.
The spectrum is finally dominated by the high-energy tail ($A_1 <0.1$) in the region $U_{fc}\gtrsim 0.7$ for $M=5$.

The energy scale $\Delta_{1}$ is equal to $\Delta_0=0.063$ at $U_{fc}=0$, and increases with increasing $U_{fc}$ in the weak-coupling regime.
In the strong-coupling regime, on the other hand, $\Delta_{1}$ turns to decrease for $M>1$. In particular, an extrapolation of the data for $M=5$ to larger $U_{fc}$ suggests an existence of a critical point characterized by $\Delta_1=0$ around $U_{fc}^* \approx 0.9$. However, we could not reach the critical point because of a numerical difficulty in the strong-coupling regime. 
The difficulty arises not only from the increased computational 
time with increasing $M$, but also from a decrease of the acceptance rate of the Monte Carlo updates,
which is  similar to the case of the ordinary Anderson model with strong Coulomb repulsion.

The non-monotonic behavior of $\Delta_1$ is a consequence of the competition 
between the exciton effect coming from the hybridizing channel 
and the Anderson orthogonality effect from the screening channels as discussed in Section~\ref{sec-analytic_approach}.
Since there are no additional  screening channels in the case of $M=1$, the low energy scale $\Delta_{1}$ increases monotonically against $U_{fc}$.
In contrast to $\Delta_1$, the width $\Delta_{2}$ of the additional peak (energy scale of the high-energy tail) does not show distinct $M$ dependence
at least for $U_{fc}<0.5$.

\begin{figure}
\centering
\includegraphics[width=0.95\hsize]{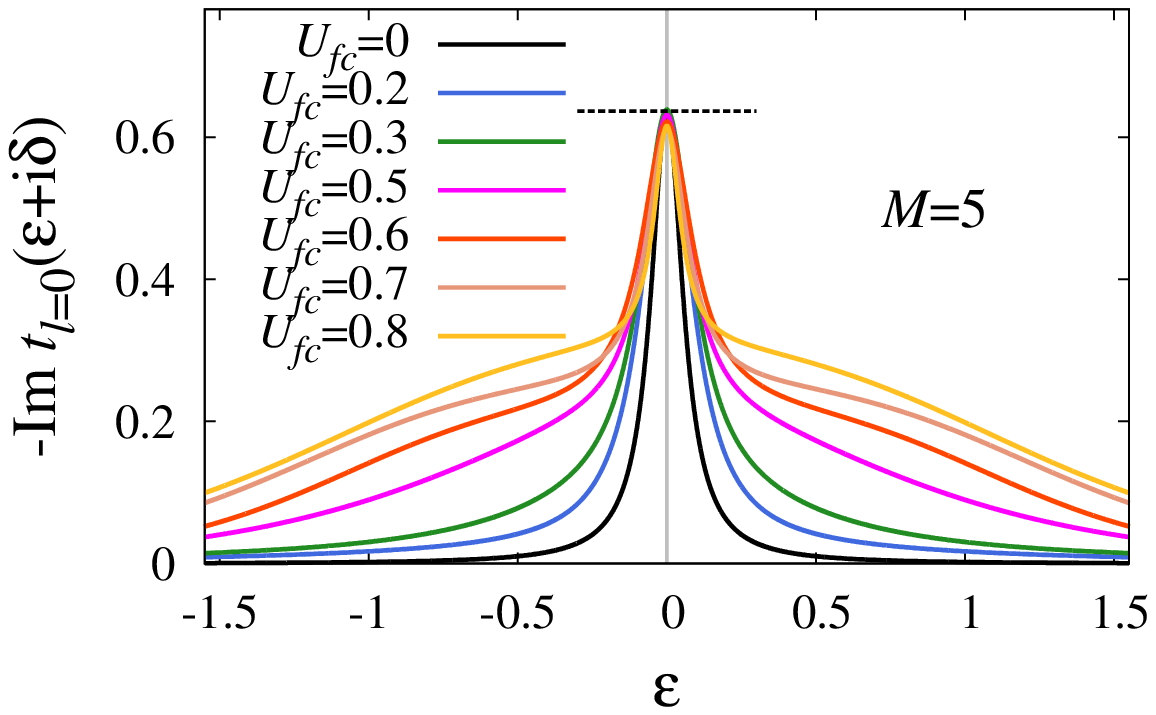}
\includegraphics[width=0.95\hsize]{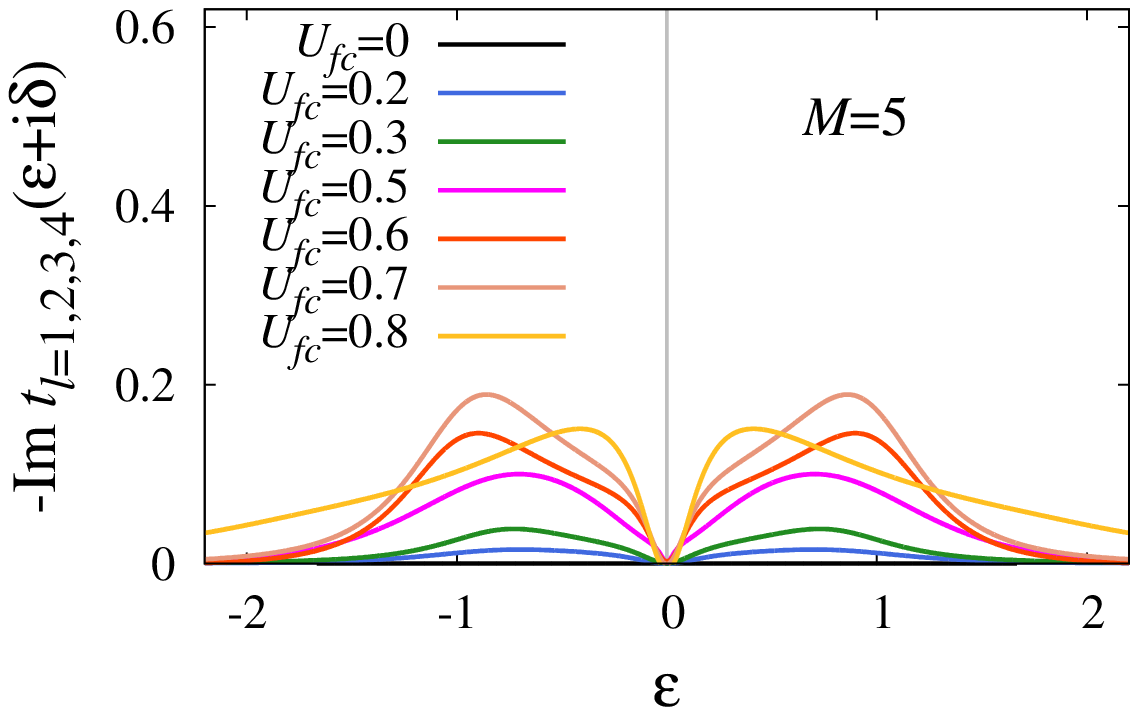}
\caption{(Color online) Energy dependence of the imaginary part of the $t$-matrix for channel number $M=5$ for $\ell=0$ ({\sl top}) and $\ell\ne 0$ ({\sl bottom}). {\sl Dashed line} at zero energy in the left panel indicates Friedel's sum rule as $1/(\pi \rho_{0})$. The parameter values are chosen as $V=0.2$, $\beta=200$.}
\label{fig-tdos}
\end{figure}

\subsection{$t$-matrix and transport}\label{sec-tmatrix}

The $t$-matrix includes all informations about the scattering of the conduction electrons by the local charge.
Figure~\ref{fig-tdos} shows the $U_{fc}$-dependence of $-{\rm Im}t_{\ell} (\varepsilon+i\delta)$ with $M=5$.
We find that ${\rm Im}\, t_{\ell}(\varepsilon)$ for the scattering channel, $\ell \ne 0$, vanishes at the Fermi level ($\varepsilon=\varepsilon_{f} =0$), while that for the hybridizing channel, $\ell=0$, satisfies the Friedel's sum rule, ${\rm Im}\,t_{\ell=0}(\varepsilon = 0) = -1/(\pi \rho_{0})$
for arbitrary values of $U_{fc}$ and $M$.
This means that the phase shift $\delta$ for the resonant scattering is $\pi/2$.

The conduction electron density of states is expressed as
\begin{align}
\rho_{c, \ell}(z) &= -\frac{1}{\pi} {\rm Im}\, G_{c,\ell}(z)\nonumber\\
 &=  -\frac{1}{\pi} {\rm Im}\,  [g(z) + g(z) t_{\ell}(z) g(z)]
\end{align}
through the $t$-matrix.
Using the properties $t_{\ell=0}(0)=-i/(\pi \rho_{0})$ and $t_{\ell \ne 0}(0)=0$ together with the approximation $g(0)=-i\pi \rho_{0}$, we find that $\rho_{c,\ell=0}(0)=0$, i.e. the local density of states of conduction electrons vanishes at the Fermi level for the hybridizing channel, while $\rho_{c,\ell \ne 0}(0)=\rho_{0}$, i.e. $\rho_{c}$ is unchanged for the screening channels.

\begin{figure}
\centering
\includegraphics[width=0.95\hsize]{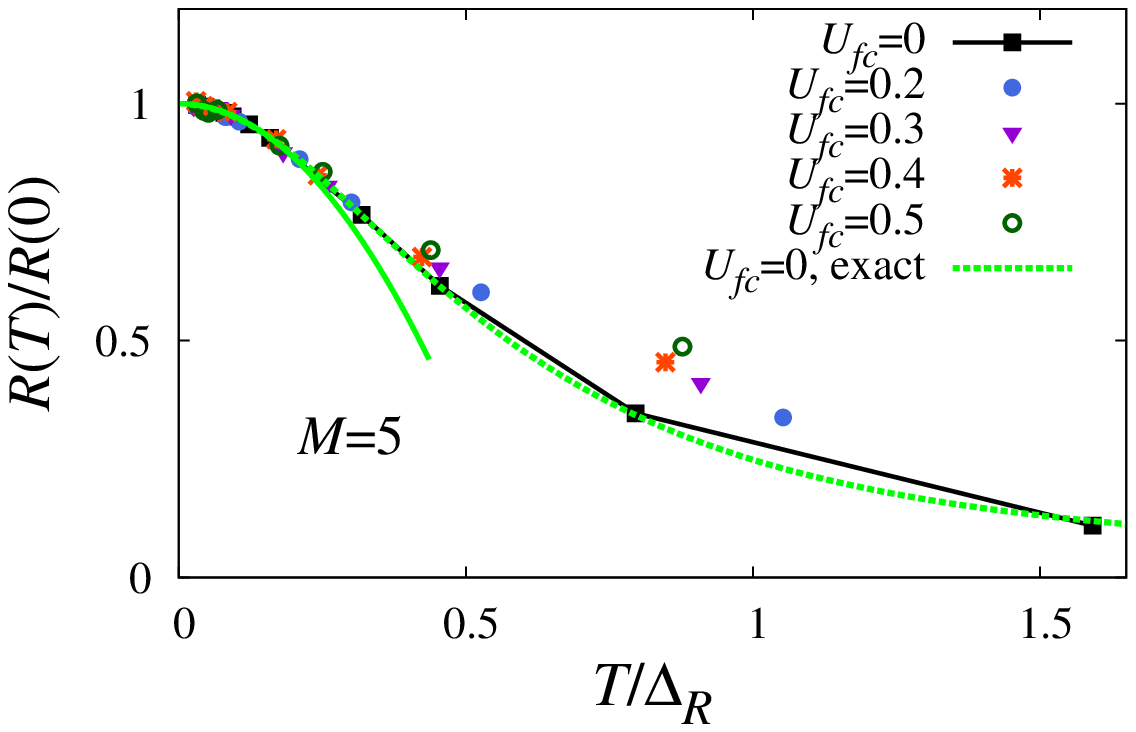}
\includegraphics[width=0.95\hsize]{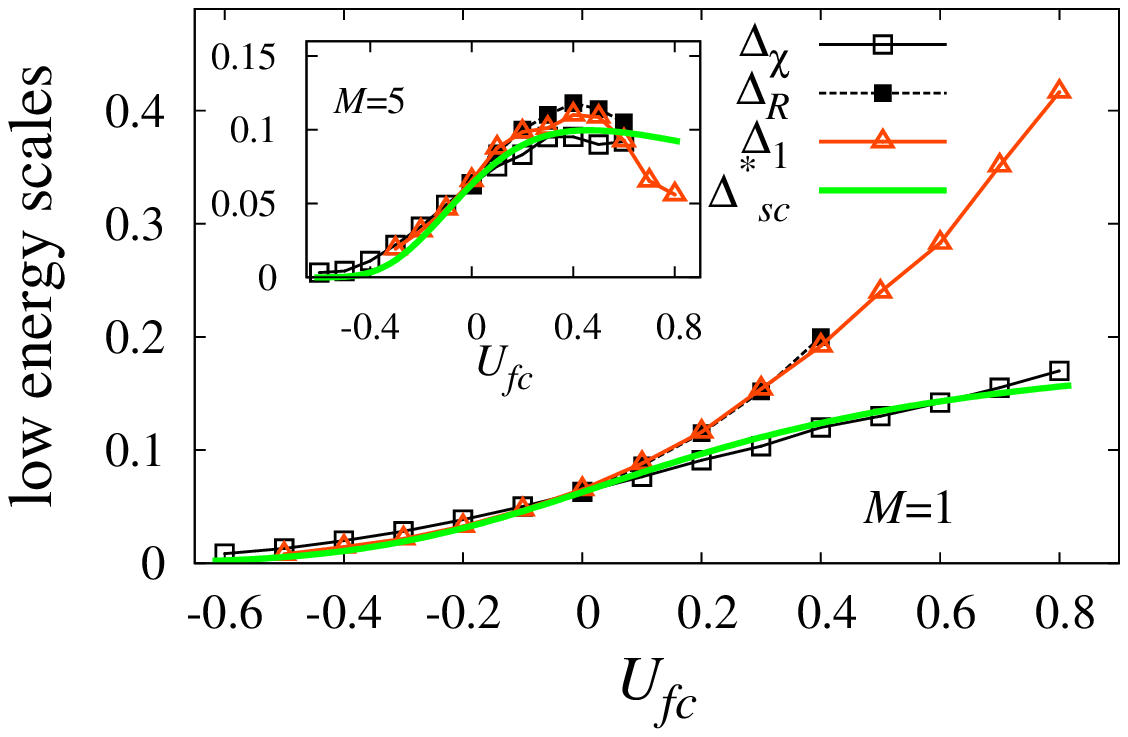}
\caption{(Color online) {\sl Top:} Electrical resistivity for channel number $M=5$ as a function of temperature for different values of $U_{fc}$.
{\sl Bottom:} Low energy scales $\Delta_{1}$,  $\Delta_{\chi}$, and $\Delta_{\rm R}$ together with the scaling result $\Delta^{\ast}_{sc}$ given by Eq.~(\ref{eq-schomd})  for $M=1$ ({\sl main panel}) and $M=5$ ({\sl inset}).
}
\label{fig-transport}
\end{figure}

To obtain more information about the low-energy dynamics, we calculate the electrical resistivity $R(T)$ at arbitrary temperature as
\begin{eqnarray}
R(T)^{-1} = \int_{-\infty}^{\infty} d \varepsilon \left[ - \frac{\partial f(\varepsilon)}{\partial \varepsilon} \right] \tau(\varepsilon)
\label{eq-resdef}
\end{eqnarray}
in the Boltzmann equation approach, where the relaxation time $\tau(\varepsilon)$ is obtained from the $t$-matrix in the hybridizing channel\cite{note-2} as $\tau(\varepsilon)^{-1} = -2 {\rm Im}\,t_{\ell=0}(\varepsilon+i\delta)$.
The temperature dependence of $R(T)$ for $M=5$ is shown in the top part of Fig.~\ref{fig-transport} 
for several values of $U_{fc}$.
Here, the temperature is scaled by a characteristic energy $\Delta_{R}$ (see below for detail).
As an accuracy check of the numerical data, we also show the analytically derived resistivity for $U_{fc}=0$
with $t_{\ell=0}(z)=V^2G_{f}(z)$. 
We confirm from the numerical data that the scaled resistivity $R(T)/R(0)$ is a universal function of $T/\Delta_{R}$
in the temperature range shown in this figure\cite{note-3}.

At low temperatures the resistivity behaves as
\begin{eqnarray}
R_{\rm LFL}(T) 
= R_0\left[1 - a \left( \frac{T}{\Delta_{R}} \right)^2\right]
\label{eq-resA}
\end{eqnarray}
within the local Fermi-liquid theory.
Using this expression, we have determined the energy scale $\Delta_{R}$. Here, we assumed that $R_0$ and $a$ are independent of $U_{fc}$, which can be determined from the Sommerfeld expansion of $R(T)$ as $R_0 = 2\alpha/(\pi \rho_{0})$ and $a=\alpha\pi^2/3$. 
The result obtained for $\Delta_{R}$ is shown in the bottom part of Fig.~\ref{fig-transport} for $M=1$ and $M=5$ together with the low energy scales $\Delta_{1}$ and $\Delta_{\chi}\equiv 2 \pi T [\psi^{-1}(2 \pi^2 T \chi_{0}(T))-1/2]$ of the local Green's function and charge susceptibility, respectively.
We find that the energy scale $\Delta_{R}$ coincides with  $\Delta_{1}$.
Thus we conclude that the low-energy scale for the $t$-matrix matches with the low-energy scale of the local Green's function.

\subsection{Self-energy}\label{sec-se}

\begin{figure}
\centering
\includegraphics[width=0.95\hsize]{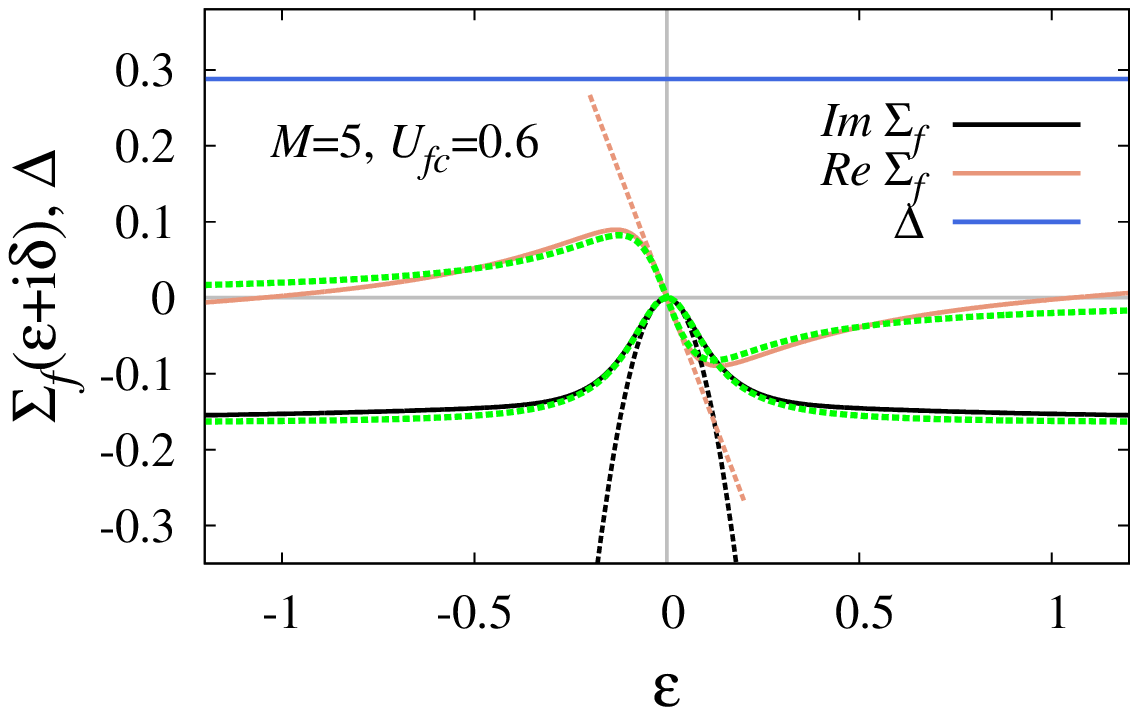}
\includegraphics[width=0.95\hsize]{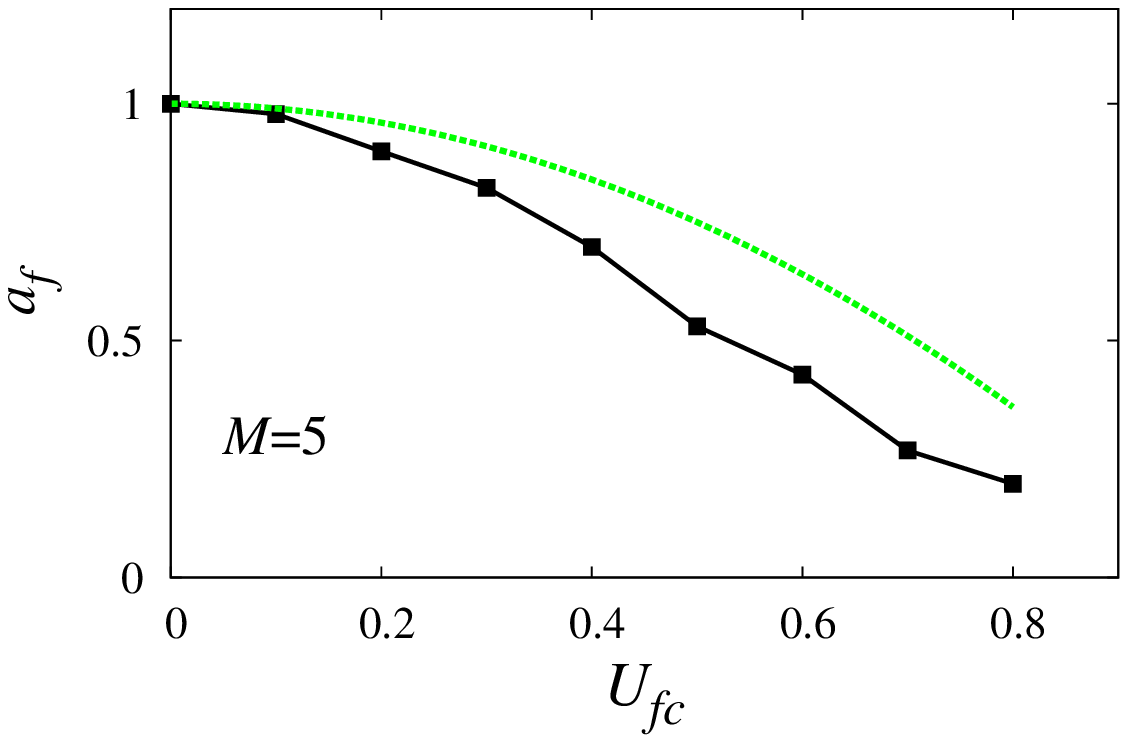}
\caption{(Color online) {\sl Top:} Energy dependence of the numerically obtained $f$-electron self-energy and parameter $\Delta$ for $M=5$ at $U_{fc}=0.6$.
The parameter values are chosen as $V=0.2$, $\beta=200$. {\sl Dashed lines} in the low-energy range correspond to the quasi-particle results to be given later in Eqs.~(\ref{eq-imSfquasi}) and (\ref{eq-reSfquasi}), while {\sl dashed green lines} show the approximation given in Eq.~(\ref{eq-sewhole}).
{\sl Bottom:} Wave-function renormalization factor $a_{f}$ for $M=5$ as a function of $U_{fc}$. {\sl Dashed green line} corresponds to the quasi-particle perturbation result $a_{f}=1-(u/u_{c})^2$ to be given in Eq.~(\ref{eqq-af2}).
}
\label{fig-selfenergy}
\end{figure}

We now discuss the self-energy to get more information on the peculiar spectra of $G_f(\varepsilon)$.
Top part of Fig.~\ref{fig-selfenergy} shows the energy dependence of the numerically obtained local self-energy $\Sigma_{f}(\varepsilon)$ together with 
$\Delta(\varepsilon) = \pi \rho_{0} |V + \Sigma_{fc}(\varepsilon)|^2$
for $M=5$ under finite $U_{fc}$.
While $\Sigma_{f}$ shows substantial energy dependence, the quantity $\Delta$ involving the self-energy $\Sigma_{fc}$ is almost independent of energy.
A crucial difference with the ordinary Fermi-liquid behavior is shown in the high-energy region, ${\rm Im}\Sigma_f(\varepsilon) \sim {\rm const.}$ It is related to the high-energy tail of ${\rm Im} G_f(\varepsilon)$.

We find that the numerical data for ${\rm Im} \Sigma_{f}(\varepsilon)$ is well approximated by the formula
\begin{eqnarray}
{\rm Im}\, \Sigma_{f}^{\rm (approx)}(\varepsilon) = -\frac{ \varepsilon^2 \Gamma}{\varepsilon^2 + \Delta_{s}^2},
\label{eq-imsf00}
\end{eqnarray}
where we use the convention that $\Gamma, \Delta_{s}>0$.
The Kramers-Kr\"onig relation constrains the 
complex self-energy to be
\begin{eqnarray}
\Sigma_{f}^{\rm (approx)}(\varepsilon) = -i\frac{\varepsilon \Gamma }{\varepsilon + i \Delta_{s}}.
\label{eq-sewhole}
\end{eqnarray}
Fit with the approximation~(\ref{eq-sewhole}) for both the imaginary and real parts of $\Sigma_{f}$ is shown in top part of Fig.~\ref{fig-selfenergy} by {\sl dashed green line}.
Formula~(\ref{eq-sewhole}) reproduces the Fermi-liquid properties ${\rm Im}\, \Sigma_{f}(\varepsilon) \propto - \varepsilon^2$ and ${\rm Re}\, \Sigma_{f}(\varepsilon) \propto - \varepsilon$ in the low-energy range $|\varepsilon| \lesssim {\cal O}( \Delta_{s})$. 
We note that the behavior ${\rm Im}\, \Sigma_{f}(\varepsilon)\sim$ const. at large energies
should break down for $\varepsilon\gg D$. The apparent constant behavior suggests the presence of an additional characteristic energy much larger than $\Delta_{s}$.

The $U_{fc}$-dependence of the wave-function renormalization factor $a_{f}$ for $M=5$ is shown in the bottom part of Fig.~\ref{fig-selfenergy}, which we obtain from the numerical self-energy data as
\begin{eqnarray}
a_{f} = \left[1 - {\rm Im}\, \Sigma_{f}(i\varepsilon_{0})/\varepsilon_{0} \right]^{-1}.
\label{eqq-afnum}
\end{eqnarray}
The decreasing feature of $a_{f}$ with $M>1$ as the Coulomb interaction is increased indicates the formation of a correlated Fermi-liquid state with large quasi-particle effective mass $m^{\ast}$ since $m^{\ast}\sim 1/a_{f}$.
Vanishing of $a_{f}$ indicates the quantum critical point where the local charge decouples from the conduction electrons. Unfortunately, this critical point is difficult to be reached by the continuous-time quantum Monte Carlo method because of the numerical difficulties mentioned in Section~\ref{sec-gf}.

\subsection{Relation between energy scales}

So far, we derived two sets of energy scales from the numerical data: $\{\Delta_{1}, \Delta_{2}\}$ from the Green's function, and $\{\Gamma, \Delta_{s}\}$ from the self-energy.
Those are related with each other.
In the following, we derive the formula connecting them.

From Eq.~(\ref{eq-Gfdef}), the $f$-electron Green's function $G_f(\varepsilon)$ is given by
\begin{eqnarray}
G_{f}(\varepsilon) = \frac{1}{\varepsilon + i\Delta - \Sigma_{f}(\varepsilon)}.
\label{eq-Gfquasi0}
\end{eqnarray}
Here, $\Delta$ denotes an effective hybridization strength defined by $\Delta = \pi \rho_{0} |V+\Sigma_{fc}(0)|^2$.
We neglected $\Sigma_{c,\ell=0}$ and the energy dependence of $\Sigma_{fc}(\varepsilon)$ according to the numerical results shown in Fig.~\ref{fig-selfenergy}. 
By replacing $\Sigma_{f}$ with $\Sigma_{f}^{\rm (approx)}$ given by Eq.~(\ref{eq-sewhole}), and equating $G_{f}$ to the two-Lorentzian form $G_{f}^{\rm (approx)}$ given by Eq.~(\ref{eq-zfdosLorentz}), 
we obtain the following relations:  
\begin{align}
A_{1}\Delta_{2} + (1-A_{1})\Delta_{1} &= \Delta_{s}, \label{eq-1}\\
\Delta_{1} \Delta_{2} &= \Delta \Delta_{s},\label{eq-2} \\
\Delta_{1} + \Delta_{2} &= \Delta + \Delta_{s} + \Gamma,\label{eq-3}\\
A_{2} &= 1 - A_{1}.\label{eq-4}
\end{align}
Solving Eqs.~(\ref{eq-1})-(\ref{eq-4}) for $\Delta_{1}$ and $\Delta_{2}$ we obtain 
\begin{align}
\Delta_{1,2} &= \frac{1}{2} \left[ (\Gamma + \Delta + \Delta_{s})
\right. \nonumber\\
&\mp \left. \sqrt{(\Gamma + \Delta + \Delta_{s})^2 - 4\Delta \Delta_{s} }  \right].
\label{Delta_12}
\end{align}

\begin{figure}
\centering
\includegraphics[width=0.95\hsize]{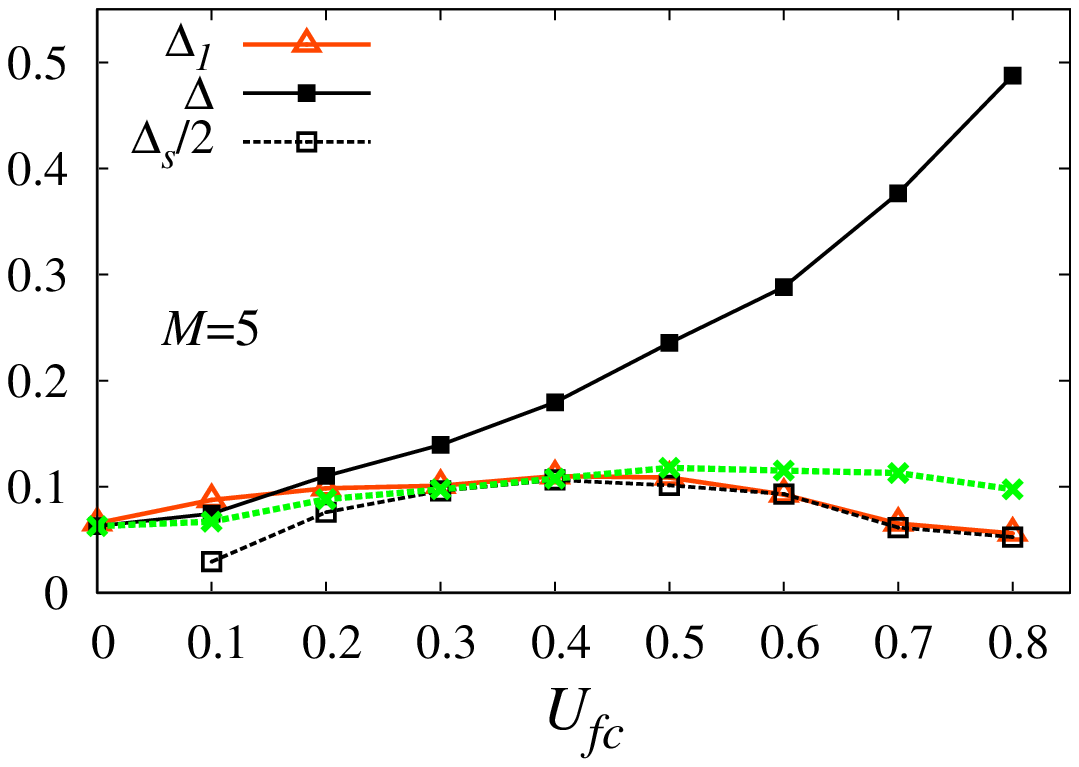}
\includegraphics[width=0.95\hsize]{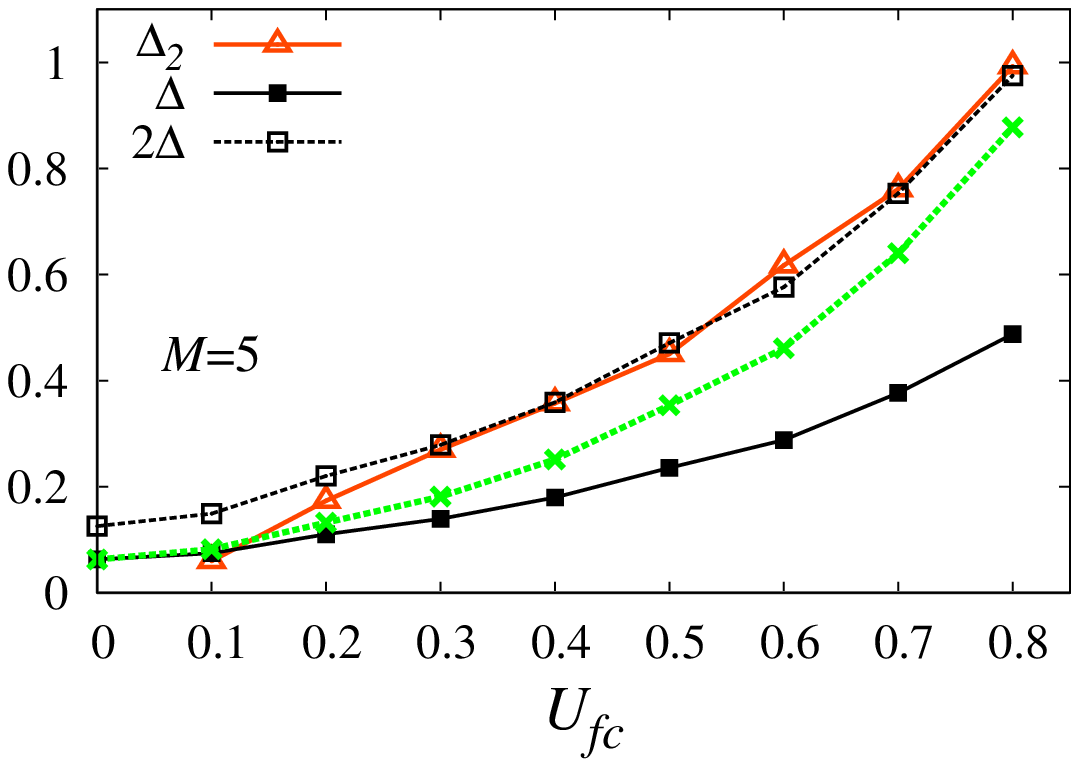}
\caption{(Color online) Energy scales $\Delta_{1}$ ({\sl top}) and $\Delta_{2}$ ({\sl bottom}) together with the energy scales $\Delta_{s}$ and  $\Delta$ as a function of $U_{fc}$ for $M=5$. 
The parameter values are chosen as $V=0.2$, $\beta=200$.
{\sl Dashed green lines} correspond to the quasi-particle perturbation result  to be given later in Eq.~(\ref{eq-d1d2fpt}).
}
\label{fig-numcheck}
\end{figure}

Now we compare $\{\Delta_{1}, \Delta_{2}\}$ with other energy scales such as $\Delta$.
Figure~\ref{fig-numcheck} shows their $U_{fc}$ dependence for $M=5$.
The enhancement of $\Delta$ over the bare hybridization is due to the exciton effect. 
Concerning 
$\Delta_{1,2}$, we find that\\
(i) $\Delta_{1}$ shows a crossover from $\Delta$ to $\Delta_{s}/2$ with increasing values of $U_{fc}$, while \\
(ii) $\Delta_{2}$ interpolates between $\Delta$ and $2\Delta$.\\
We will discuss these characteristics 
by a microscopic Fermi-liquid theory in the next section.

\section{Quasi-Particle Perturbation Theory}\label{sec-discussion}

\subsection{Second-order self-energy}

For descriptions of low-energy properties, we may work with a quasi-particle Green's function.
Assuming the ordinary Fermi-liquid properties for $\Sigma_f(\varepsilon)$ in Eq.~(\ref{eq-Gfquasi0}), we obtain
\begin{eqnarray}
G^{\ast}_{f}(\varepsilon) = \frac{1}{\varepsilon + i\Delta^{\ast} - \Sigma^{\ast}_{f}(\varepsilon)},
\label{eq-Gfquasi}
\end{eqnarray}
where $G_{f} = a_{f}G_{f}^{\ast}$, $\Delta^{\ast} = a_{f} \Delta$, and $\Sigma^{\ast}_{f} = a_{f} \Sigma_{f}$
with 
\begin{eqnarray}
a_{f} = \left[1 - \partial \Sigma_{f}(\varepsilon) / \partial \varepsilon \right]_{\varepsilon=0}^{-1},
\label{eqq-af}
\end{eqnarray}
being the wave-function renormalization factor.
Since $a_f<1$, $G^{\ast}_{f}$ has a smaller characteristic energy scale, $\Delta^{\ast} < \Delta$,
and quasi-particle descriptions are valid in the region $| \varepsilon  | \lesssim {\cal O}(\Delta^{\ast})$.

\begin{figure}
\centering
\includegraphics[width=0.70\hsize]{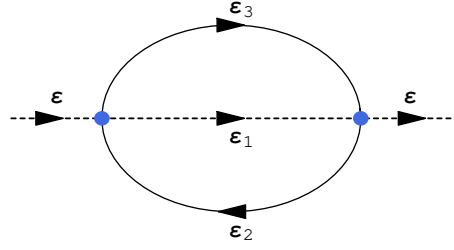}
\caption{(Color online) Second-order self-energy diagram with {\sl dashed line} corresponding to local ($f$) state and {\sl solid line} representing conduction electron state. {\sl Dots} represent the Coulomb interaction $U_{fc}$.}
\label{fig-diagram}
\end{figure}

We now derive $\Sigma_{f}^{\ast}$ by a perturbation theory with respect to $U_{fc}$. 
The leading contribution is given by the second-order diagram shown in Fig.~\ref{fig-diagram}.
The explicit expression reads
\begin{align}
\Sigma_{f}^{\ast}(i\varepsilon_n)
&= -u^2 \frac{1}{\beta^2} \sum_{n_1 n_2}
G^{\ast}_f(i\varepsilon_{n_1}) \sum_{\ell} \bar{G}_{c,\ell}(i\varepsilon_{n_2})
\nonumber\\
\times& \bar{G}_{c,\ell}(i\varepsilon_{n}-i\varepsilon_{n_1}+i\varepsilon_{n_2}),
\label{eq-Sfquasi}
\end{align}
where
$\bar{G}_{c,\ell}$ is the local component of conduction-electron Green's function renormalized by the hybridization process.
The important point here is that the influence of $U_{fc}$ in the hybridizing channel $\ell=0$ is already included into $G_f^{\ast}$ as the effective hybridization $\Delta$.
Since the local density of states $\rho_{c, \ell=0}(0)$ vanishes at the the Fermi level as presented in Sec.~\ref{sec-tmatrix},
the channels contributing to $
- {\rm Im}\, \Sigma_{f}^{\ast} (\varepsilon) $
are those with $\ell\neq 0$.
The Green's functions $\bar{G}_{c,\ell \neq0}$ of these non-hybridizing channels
may be replaced by $\bar{g}_{\bm{k}}(i\varepsilon) = (1/N)\sum_{\bm{k}} (i\varepsilon - \varepsilon_{\bm{k}})^{-1}$.
Performing the Matsubara summations and taking the imaginary part, we obtain 
\begin{align}
- {\rm Im}\, \Sigma_{f}^{\ast} (\varepsilon) &= u^2 (M-1) \int_{0}^{\infty} d \varepsilon_{1} \int_{-\infty}^{0} d \varepsilon_{2} \int_{0}^{\infty }d \varepsilon_{3}
\nonumber\\
 \times& 2 \pi \rho_{f}^{\ast} (\varepsilon_{1}) \delta(\varepsilon - \varepsilon_{1} + \varepsilon_{2} - \varepsilon_{3}),
\label{eq-imSfquasi2}
\end{align}
where $\rho_{f}^{\ast}=-(1/\pi) {\rm Im}G_f^{\ast}$.
In the low-energy limit, $\varepsilon \ll  \Delta^{\ast}$, 
we may replace $G^{\ast}_f(i\varepsilon)$ by 
$G^{\ast 0 }_f(i\varepsilon)
 = (i\varepsilon-i\Delta^{\ast})^{-1}$,
and 
the integrals can be performed analytically to yield
\begin{eqnarray}
- {\rm Im}\, \Sigma_{f}^{\ast}(\varepsilon) = u^2 (M-1) 
\varepsilon^2/\Delta^{\ast}
\label{eq-imSfquasi}
\end{eqnarray}
with neglect of terms of order $\Delta^*(\varepsilon/{\Delta^*})^4$.

The real part of ${\Sigma^{\ast}_{f}}$ may be evaluated using Eq.~(\ref{eq-Sfquasi}) as well. However, in contrast to the imaginary part, 
high-energy processes give finite contributions in this case.
Then we cannot use the low-energy form for $G_f^{\ast}$.
We instead derive the low-energy limit of ${\rm Re}{\Sigma^{\ast}_{f}}$ by the following argument.
The particle-hole symmetry ensures the condition ${\rm Re}{\Sigma^{\ast}_{f}}(0)=0$.
It follows that the low-energy expression for ${\rm Re}{\Sigma^{\ast}_{f}}$ may be given within ${\cal O}(u^2)$ as
\begin{align}
{\rm Re}{\Sigma^{\ast}_{f}}(\varepsilon) &= - 
(u/u_c)^2 \varepsilon,
\label{eq-reSfquasi}
\end{align}
where $u_c$ is the critical value of $\rho_0U_{fc}$ that gives the quantum critical point.
According to the second-order scaling we obtain Eq.~(\ref{eq-upm}) which reduces to
$u_c^2 \rightarrow 1/M$ in the limit of 
$M\gg 1$. 
Since $u_c^2 \ll 1$ for large $M$, the second-order self-energy should give precise account for $u<u_c$.
We then extrapolate the second-order theory for smaller $M$ by 
modifying the form to $u_c^2 = 1/(M-1)$, which avoids correctly the critical point at $M=1$.
Combining with Eq.(\ref{eqq-af}), we then obtain
\begin{eqnarray}
a_{f} =  1 - (M-1) u^2 = 1 - \left( u/u_{c}\right)^2, 
\label{eqq-af2}
\end{eqnarray}
which is reduced to zero at $u=u_{c}=(M-1)^{-1/2}$. 
Finally, the quasi-particle Green's function is obtained from Eq.~(\ref{eq-Gfquasi}) as
\begin{eqnarray}
G_{f}^{\ast}(\varepsilon) = \left\{ \varepsilon + i\Delta^{\ast} 
+ i  (1-a_{f}) \varepsilon^2/\Delta^{\ast} 
\right\}^{-1}
\label{eq-quasiGf2} 
\end{eqnarray}
by using Eqs.~(\ref{eq-imSfquasi}), (\ref{eq-reSfquasi}), and (\ref{eqq-af2}).

\subsection{Comparison with numerical data}

Comparing the approximate formula~(\ref{eq-sewhole}) of the self-energy with Eqs.~(\ref{eq-imSfquasi}) and (\ref{eq-reSfquasi}), we obtain the following correspondences between the phenomenology and quasi-particle theory:
\begin{align}
- \left.\frac{\partial {\rm Re}\, \Sigma_{f}(\varepsilon)}{\partial \varepsilon}\right|_{\varepsilon=0} &= \Gamma/\Delta_{s} 
= a_{f}^{-1} -1,
\label{eq-corresp1}\\
- \frac{1}{2} \left.\frac{\partial^2 {\rm Im}\, \Sigma_{f}(\varepsilon)}{\partial \varepsilon^2}\right|_{\varepsilon=0} &= \Gamma/\Delta_{s}^2 
= (a_{f}^{-1} -1)/\Delta^*,
\label{eq-corresp2}
\end{align}
which give the relations
\begin{align}
\Delta_{s} &=  \Delta^{\ast}= a_{f} \Delta,\label{eq-deltas0}\\
\Gamma &= (1-a_{f}) \Delta = (M-1) u^2 \Delta.\label{eq-gamma0}
\end{align}
Thus, the phenomenological parameters $\Gamma$  and $\Delta_{s}$ in the approximate self-energy are determined by the 
single parameter $a_{f}$ for given $\Delta$.
This is a strong constraint imposed by the quasi-particle Fermi-liquid theory.
It is thus interesting to check the accuracy of the quasi-particle perturbation theory in the light of the numerical results. 

First we examine the relation given in Eq.~(\ref{eqq-af2}) which expresses the wave-function renormalization factor $a_{f}$ in a simple way. The $U_{fc}$-dependence of the estimate $1-(u/u_c)^2=1 - (M-1) u^2$ with $u=\rho_{0}U_{fc}$ is shown as {\sl green dashed line}  in the bottom part of Fig.~\ref{fig-selfenergy} together with the numerically obtained  $a_{f}$. We can observe that the simple, second-order quasi-particle formula describes the wave-function renormalization factor reasonably well.

Next we discuss the case of self-energy.
Top part of Fig.~\ref{fig-selfenergy} includes also the fit of the numerical self-energy by the result of the quasi-particle theory given in  Eqs.~(\ref{eq-imSfquasi}) and (\ref{eq-reSfquasi}) as {\sl dashed lines}.
We conclude that the theory works well in the Fermi-liquid range $|\varepsilon | \lesssim \Delta^{\ast}$, i.e. at low energies.

Using the relations~(\ref{eq-deltas0}) and (\ref{eq-gamma0}), the
approximate self-energy given in Eq.~(\ref{eq-sewhole}) is expressed as
\begin{eqnarray}
\Sigma^{\rm (approx)}_{f} (\varepsilon) = -i \frac{(1-a_{f})\Delta \varepsilon}{\varepsilon + i a_{f} \Delta}
\label{eq-se10}
\end{eqnarray}
in the quasi-particle theory.
This formula is also shown in the top part of  Fig.~\ref{fig-selfenergy} by {\sl green lines}, and 
gives an excellent fit of the numerical self-energy in the whole energy range.
This is not surprising since the approximation given in Eq.~(\ref{eq-sewhole}) has only two parameters, $\Delta_{s}$ and $\Gamma$, and if the quasi-particle theory fits the self-energy around $\varepsilon \sim 0$, it will fit also the curve in the whole energy range.

Finally, we discuss the energy scales $\Delta_{1}$ and $\Delta_{2}$.
Namely, we obtain from Eq.(\ref{Delta_12}) the following relation:
\begin{eqnarray}
\Delta_{1,2}/\Delta = 
1\mp \sqrt{1-a_f} =
1\mp u/u_c
\label{eq-d1d2fpt}
\end{eqnarray}
by using Eqs.~(\ref{eqq-af2}), (\ref{eq-deltas0}), and (\ref{eq-gamma0}).
The limiting behavior is thus given by
\begin{align}
\Delta_{1}/\Delta = 1 - \sqrt{1-a_{f}}  = 
\begin{cases}
a_{f}/2 = \Delta_{s}/(2\Delta) & \text{if $a_{f} \ll 1$} \\
1 & \text{if $a_{f} \approx 1$}
\label{eqq-d1}
\end{cases},
\end{align} 
and
\begin{align}
\Delta_{2}/\Delta = 
1 + \sqrt{1-a_{f}}  = 
\begin{cases}
2 & \text{if $a_{f} \ll 1$} \\
1 & \text{if $a_{f} \approx 1$}
\label{eqq-d2}
\end{cases}.
\end{align}
Actually, these are exactly the limits that we found from the analysis of the numerical data (see Fig.~\ref{fig-numcheck}).
We show the expressions $\Delta(1\mp u/u_c)$ from Eq.~(\ref{eq-d1d2fpt}) in Fig.~\ref{fig-numcheck} by {\sl dashed green curves} as well, which describe $\Delta_{1}$ and $\Delta_{2}$ relatively well.
We plotted the energy scales $\Delta_{1}$ and $\Delta_{2}$ in Fig.~\ref{fig-numcheck} instead of the ratios $\Delta_{1,2}/\Delta$ because numerical errors are larger in $\Delta_{1,2}/\Delta$ in the low-$U_{fc}$ range since the fit with two Lorentzians is unambiguous in this region; the contribution from the wider Lorentzian is very small.

We conclude that the extrapolated quasi-particle theory describes the $U_{fc}$ dependence of both the energy scales $\{\Delta_{1}, \Delta_{2}\}$ and the self-energy in a surprisingly wide range of $U_{fc}$.
However, this theory does not give any microscopic origin neither for the constant behavior in ${\rm Im}\, \Sigma_{f}(\varepsilon)$ at large energies nor for the wider Lorentzian component in ${\rm Im}\, G_{f}(\varepsilon)$ since these properties appear out of the valid energy range of the perturbation theory.

\section{Summary}\label{sec-summary}

In this paper we have studied the multichannel interacting resonant level model by means of the continuous-time quantum Monte Carlo method in a wide range of the Coulomb interaction $U_{fc}$ and channel number $M$.
Thermodynamic and dynamic properties have been derived 
accurately and have been discussed in the light of analytic approaches.

We find that thermodynamics of MIRLM such as the local charge susceptibility is entirely determined by the single energy scale $\Delta_{sc}^{\ast}$ within the scaling theory.
On the other hand, dynamics 
contains multiple energy scales beyond the simple scaling theory.
For example, we find that the single-particle excitation spectrum acquires a high-energy tail for $M>1$ 
with increasing values of $U_{fc}$ in addition to the narrow resonance around the Fermi level.
This composite line shape can be 
well described by the sum of two Lorentzians, where the narrow one with energy scale $\Delta_1$ corresponds to the $f$ level peaked at $\varepsilon_f=0$, 
while the wider one with the scale $\Delta_2$ gives an account of the high-energy tail. 
We find that while the narrow energy scale $\Delta_1$ shows strong $M$-dependence and non-monotonic behavior for $M>1$, the larger energy scale $\Delta_2$ is independent of $M$ for moderate values of $U_{fc}$.

The numerically obtained self-energy also shows unusual behavior: its imaginary part is 
nearly constant in the high-energy region,
which is related to the 
tail of the Green's function.
We note that a three-peak structure similar to the case of the symmetric Anderson model is expected for ${\rm Im}\, \Sigma_{f}$ close to the quantum critical point since the renormalized hybridization 
tends to zero, and therefore 
addition or removal of an $f$ electron from the ground state 
may accompany the additional peaks.
Actually, preliminary numerical data showing this situation has already been obtained numerically\cite{miyazawa}.

A quasi-particle perturbation theory from the Fermi-liquid fixed point is used for the microscopic understanding and description of the low-energy part of the single-particle spectra.
The microscopic theory provides a constraint among the parameters in the phenomenological theory, and gives description of the spectrum by a single parameter $a_f$ for given $\Delta$.

Finally, we propose that the multichannel interacting resonant level scenario might be responsible for the peculiar heavy-fermion state of certain Samarium compounds with large mass enhancement and magnetic field insensitivity. 
Namely, we speculate that in the regime near the quantum critical points $u_{\pm}$ with vanishingly small effective hybridization a highly renormalized heavy-fermion state is developed with large effective mass, which is independent of the external magnetic field since only charge is involved.   It remains to see how large is the number of active channels in real materials.

\section*{Acknowledgment}

This work is supported by the Marie Curie Grants PIRG-GA-2010-276834 and the Hungarian Scientific Research Funds No. K106047. A. K. acknowledges the Bolyai Program of the Hungarian Academy of Sciences.

\section*{Appendix A: Renormalized Green's functions}\label{app-greensfunctions}

\renewcommand{\theequation}{A$\cdot$\arabic{equation}}
\setcounter{equation}{0}

The on-site Green's functions of the MIRLM is expressed in the following matrix form
\begin{align}
&
\begin{pmatrix}
\hat{G}_{c,\ell \ne 0}(z) & 0 & 0\\ 
0 & G_{c,0}(z)   & G_{cf}(z) \\
0& G_{cf}(z)^{\ast} & G_{f}(z)  
\end{pmatrix}^{-1} = \nonumber \\
&
\begin{pmatrix}
\hat{g}(z)  & 0 & 0\\ 
0 & g(z)   & 0 \\
0 & 0 & g_{f}(z)  
\end{pmatrix}^{-1}
\nonumber\\
-&
\begin{pmatrix}
\hat{\Sigma}_{c,\ell \ne 0}(z) & 0 & 0\\ 
0 &  \Sigma_{c,0}(z)   & \widetilde{V}(z) \\
0 &  \widetilde{V}^{\ast}(z) & \Sigma_{f}(z)  
\end{pmatrix}
\label{eq-greensf}
\end{align}
with $g_{f}(z)=(z-\varepsilon_{f})^{-1}$ and $g(z)=N^{-1}\sum_{\boldsymbol k}(z - \varepsilon_{\boldsymbol k})^{-1}$, and we introduced $\widetilde{V}(z)\equiv V + \Sigma_{fc}(z)$ with $V$ being the bare hybridization. 
From Eq.~(\ref{eq-greensf}) the Green's functions are obtained as
\begin{align}
G_{f}(z) &= g_{f}(z) + g_{f}(z) \widetilde{V}(z) G_{cf}(z) + g_{f}(z)\Sigma_{f}(z)G_{f}(z)\nonumber\\
 &= \left[g_{f}(z)^{-1} - \Sigma_{f}(z) - \frac{g(z) \widetilde{V}(z)^2}{1-g(z) \Sigma_{c,0}(z)} \right]^{-1},\label{eq-Gfdef}\\
G_{cf}(z) &= g(z) \widetilde{V}(z) G_{f}(z) + g(z)\Sigma_{c,0}(z)G_{cf}(z) \nonumber\\
&= \frac{ \widetilde{V}(z) G_{f}(z)}{g(z)^{-1} - \Sigma_{c,0}(z)}, \label{eq-Gcfdef}\\
G_{c,\ell}(z)  &= g(z) + g(z) \left(  \frac{ \Sigma_{c,\ell}(z)}{1-g(z) \Sigma_{c,\ell}(z)}
\right. \nonumber\\
 &+ \left. \delta_{\ell,0}\frac{\widetilde{V}(z)^2 G_{f}(z)}{(1-g(z) \Sigma_{c,\ell}(z))^2} \right)  g(z).\label{eq-tmatrdef}
\end{align}


\begin{thebibliography}{99}
\bibitem{note-2}
We note that the contribution to the resistivity from the additional $M-1$ screening channels is expected to be small in the Fermi-liquid range, since ${\rm Im}\, t_{\ell\ne 0}(\varepsilon+i\delta)$ vanishes at the Fermi level.
\bibitem{note-3}
At higher temperatures, the numerical data deviates from the universal behavior, which might be caused by the increasing dominance of $\Delta_{2}$ in the $t$-matrix.
\end{thebibliography}
\end{document}